\documentclass[12pt,preprint]{aastex}

\usepackage{graphicx}
\usepackage{natbib}

\shorttitle{A kink-unstable filament eruption}
\shortauthors{Williams et al.}

\begin{document}

\title{Eruption of a kink-unstable filament in Active Region NOAA~10696}
\author{David~R.~Williams\altaffilmark{1}, Tibor~T\"or\"ok\altaffilmark{1}, Pascal D\'emoulin\altaffilmark{2}, Lidia~van~Driel-Gesztelyi\altaffilmark{1,2,3} and Bernhard~Kliem\altaffilmark{4}}
\altaffiltext{1}{Mullard Space Science Laboratory, University College London, Holmbury~St.~Mary, Surrey, RH5 6NT, U.K.\email{drw@mssl.ucl.ac.uk, tt@mssl.ucl.ac.uk, lvdg@mssl.ucl.ac.uk}}
\altaffiltext{2}{Observatoire de Paris, LESIA, FRE 2461 (CNRS),\\F-92195 Meudon-Principal Cedex, France\email{pascal.demoulin@obspm.fr}}
\altaffiltext{3}{Konkoly Observatory, P.O. Box 67, H-1525 Budapest, Hungary}
\altaffiltext{4}{Astrophysikalisches Institut Potsdam, 14482 Potsdam, Germany\email{bkliem@aip.de}}

%\clearpage
%\newpage
%\vspace{12cm}
\begin{abstract}
We present rapid-cadence {\sl Transition Region And Coronal Explorer} ({\sl TRACE}) observations which show evidence of a filament eruption from active region NOAA~10696, accompanied by an X2.5 flare, on 2004 November 10. The eruptive filament, which manifests as a fast coronal mass ejection some minutes later, rises as a kinking structure with an apparently exponential growth of height within {\sl TRACE}'s field of view.  We compare the characteristics of this filament eruption with MHD numerical simulations of a kink-unstable magnetic flux rope, finding excellent qualitative agreement. We suggest that, while tether-weakening by breakout-like quadrupolar reconnection may be the release mechanism for the previously confined flux rope, the driver of the expansion is most likely the MHD helical kink instability.
\end{abstract}

%@@@@@@@@@@@@@@@@@@@@@@@@@@@@@@@@@@@@@@@@@@@@@@@@@@@@@@@@@@@@@@@@@@@@@@@@@@@@@@@@@@@@@@@@@@@@@@@@

\section{Introduction}
\label{Introduction}

Elucidating the release and driver mechanisms of both coronal mass ejections 
(CMEs) and filament eruptions is a prominent issue in current solar physics research. 
Several physical mechanisms are in competition. In the following, we briefly describe 
three of them, selected because the data we analyze are able to impose constraints upon them.
   
The anchorage of magnetic field in the photosphere (and below) provides a strong stabilizing mechanism against the magnetic stresses which build magnetic pressure. For example, a stressed core field receives a strong stabilizing force from the magnetic tension of its overlying magnetic arcade, whose field lines are analogous to tethers. Cutting these tethers, via magnetic reconnection, is one possible way of allowing the core field to erupt. Starting with a simple, bipolar, sheared magnetic arcade, \cite{1989ApJ...343..971V} proposed that converging footpoint motions will force magnetic reconnection between the arcade field lines, progressively transforming the arcade core into a twisted flux tube. Such reconnection weakens the photospheric anchorage and, at some point, the remaining arcade tethers are expected to be too weak to prevent the eruption of the stressed core. In this approach, magnetic reconnection is envisioned as working at low altitude, below the stressed field which will later erupt. This approach has been developed, and compared with observations, by \cite{mooreroumel}.

An alternative to this approach is one where reconnection above the stressed 
field is responsible for the removal of overlying tethers, namely the 
magnetic ``breakout'' model (\citealt*{1999ApJ...510..485A}; \citealt{2004ApJ...617..589L}). 
This requires a large-scale, quadrupolar magnetic field configuration.  The core of the central arcade 
is assumed to be increasingly sheared by photospheric motions parallel to the magnetic polarity inversion line (IL). This core then expands with a nearly horizontal current sheet being formed around the X-point of the quadrupolar configuration.  Magnetic reconnection, at this current sheet, progressively removes the field lines above the central, sheared arcade by transferring these magnetic connections to the lateral arcades (the two side-lobes of the quadrupolar configuration).  At some point in this evolution, the feedback between the core expansion and the removal of the field lines above becomes explosive and the eruption of the core field follows.  \cite{2004ApJ...611..545G} analyzed an eruptive flare which has all the main characteristics of 
the breakout model.  A lateral version of this breakout model is also possible: here the most stressed field is in one of the side-lobes of the quadrupolar arrangement.  In this case, reconnection only weakens the stabilizing upper arcade (by transferring it to larger scales).  Observed flares demonstrating the characteristics of this mechanism have been analyzed by \cite{2000ApJ...540.1126A}, \cite{2004ApJ...611..545G} and \cite{2004ApJ...613.1221S}.

Another possible eruption mechanism is based on the helical kink 
instability (KI) of a coronal magnetic flux rope.  The KI occurs if the twist, a measure of the number of windings of the field lines about the rope axis, exceeds a critical value \citep{1981GApFD..17..297H}, leading to a helical deformation of the flux rope axis.  In coronal applications of curved and line-tied flux-rope models, this deformation results in an exponentially growing height of the flux rope's apex in the linear phase of the instability, 
as shown in numerical simulations \citep*{2003ApJ...589L.105F,2004A&A...413L..27T}.  Based on these simulations,  \cite{tk05} and \cite{fan05} succeeded in modelling the ejection of a coronal flux rope from the Sun. \cite{2005ApJ...622L..69R} also presented observational evidence that the KI is the driver in a number of filament eruptions.

Are these CME models mutually exclusive, then, or is a combination at work on the Sun?  In this letter, we investigate a well observed event that occurred on 2004 November 10 in order to look for clues.

%@@@@@@@@@@@@@@@@@@@@@@@@@@@@@@@@@@@@@@@@@@@@@@@@@@@@@@@@@@@@@@@@@@@@@@@@@@@@@@@@@@@@@@@@@@@@@@@@
\section{Observations \& Reduction}  
\label{obsred}

%\subsection{Observations}
\label{obsall}

\begin{figure*}
\begin{center}
\includegraphics[height=6.0in,angle=90]{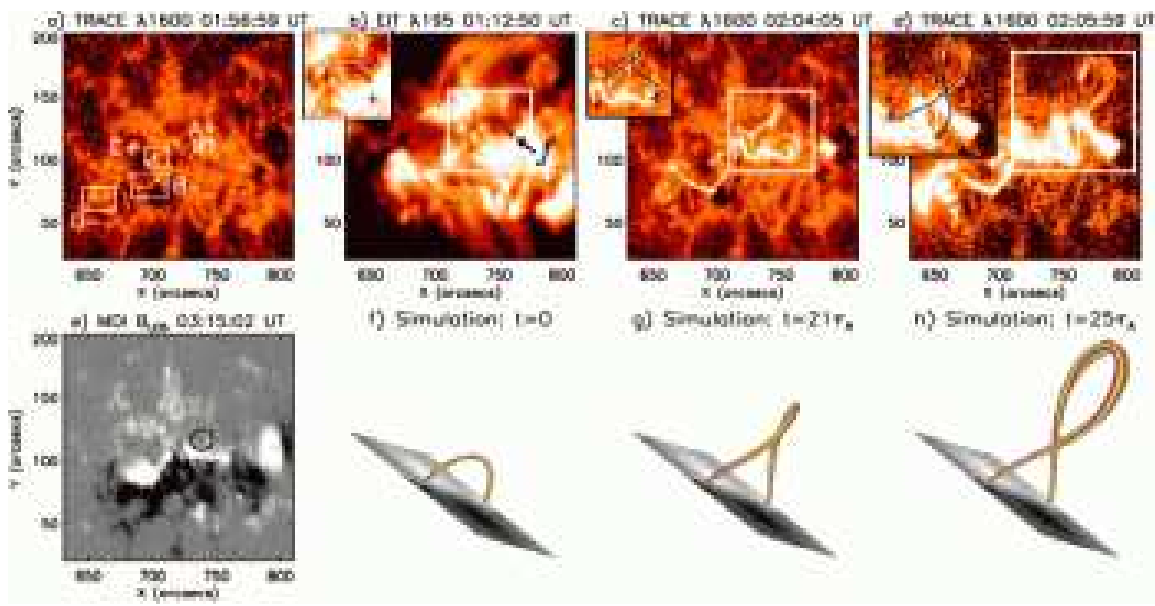}
%
%N.B. for online edition use the file f1_colour.eps instead, as follows:
%\includegraphics[width=8cm,angle=90]{f1}
\caption{a) {\sl TRACE} 1600~{\AA} image: overlaid are four boxes to show the areas of the quadripartite brightening before the impulsive phase of the flare.  b) EIT high-resolution image at 195~{\AA} showing filament $f$ in its pre-eruption position between (725\arcsec, 115\arcsec) and (755\arcsec, 110\arcsec); white box shows the area covered by the adjacent inset. Inset shows filament traced by a solid white line, with the apparent footpoints of the filament marked by black crosses. c) and d) {\sl TRACE} 1600~{\AA} images showing the evolution of the bright kinked filament. Insets as for (b) but with black line tracing the kinked shape of the erupting structure. Note the reversed-J shape of the ribbons between the filament's apparent feet in c).  e) The first MDI longitudinal magnetogram after the flare's impulsive phase. The black circle denotes a small patch of negative flux which has appeared north of the IL.  f) thru h) magnetic field lines outlining the core of the kink-unstable flux rope at $t$=0, 21 and 25~Alfv\'en times in the numerical simulation of \cite{tk05}. The grayscale levels in the plane indicate positive (white) and negative (black) concentrations of the magnetic field strength component normal to the plane. Figs. a), c) and d) form part of {\tt ktrace.mpg}, Fig. e) is part of {\tt kmdi.mpg}, and Figs. f) thru h) are taken from {\tt ksimu.mpg} in the electronic edition of the {\it Astrophysical Journal}.\label{traceimages}}
\end{center}
\end{figure*}

We differentially rotated 162 full-disk magnetograms, taken with the {\sl Solar and Heliospheric Observatory's} Michelson Doppler Imager ({\sl SoHO}/MDI; \citealt{mdi}) between 2004 November 01 and 2004 November 10, to the central meridian, and analyzed subfields centered around Region NOAA 10696. This allows the deduction of a negative sign of magnetic helicity for the flux emerging in this region \citep{lopezwrithe}. During the evolution of this region, there is continuous input of magnetic energy and helicity, which renders it prolific in flares and CMEs. Here, we have investigated the eruption of a filament in the centre of Region 10696 on 2004 November 10 (Fig.~\ref{traceimages}), which seems to be part of this longer evolution and led to a very fast CME. We note that, in the vicinity of the filament (hereafter indicated by $f$), there is a westward shearing motion of large positive-polarity elements, parallel to the IL.

The {\sl Transition Region And Coronal Explorer} ({\sl TRACE}; \citealt{trace}) observed this eruption primarily in the 1600~{\AA} band. From 01:21:56~UT, these observations were made at a cadence of $\sim$34~s until 01:57:30~UT, and thereafter at a mean cadence of 3.7~s. In order to show the pre-eruption filament, we also present an image (Fig.~\ref{traceimages}b) from the 195~{\AA} filter of {\sl SoHO}'s Extreme-ultraviolet Imaging Telescope (EIT; \citealt{eit}) at 01:12:50~UT.

For consistency, all data shown in Figs.~\ref{traceimages}a to \ref{traceimages}e (and the movie {\tt ktrace.mpg}) were differentially rotated back to the time of the last MDI image before eruption at 01:39:02~UT.

\begin{figure}
\begin{center}
\includegraphics[height=6.0in,angle=90]{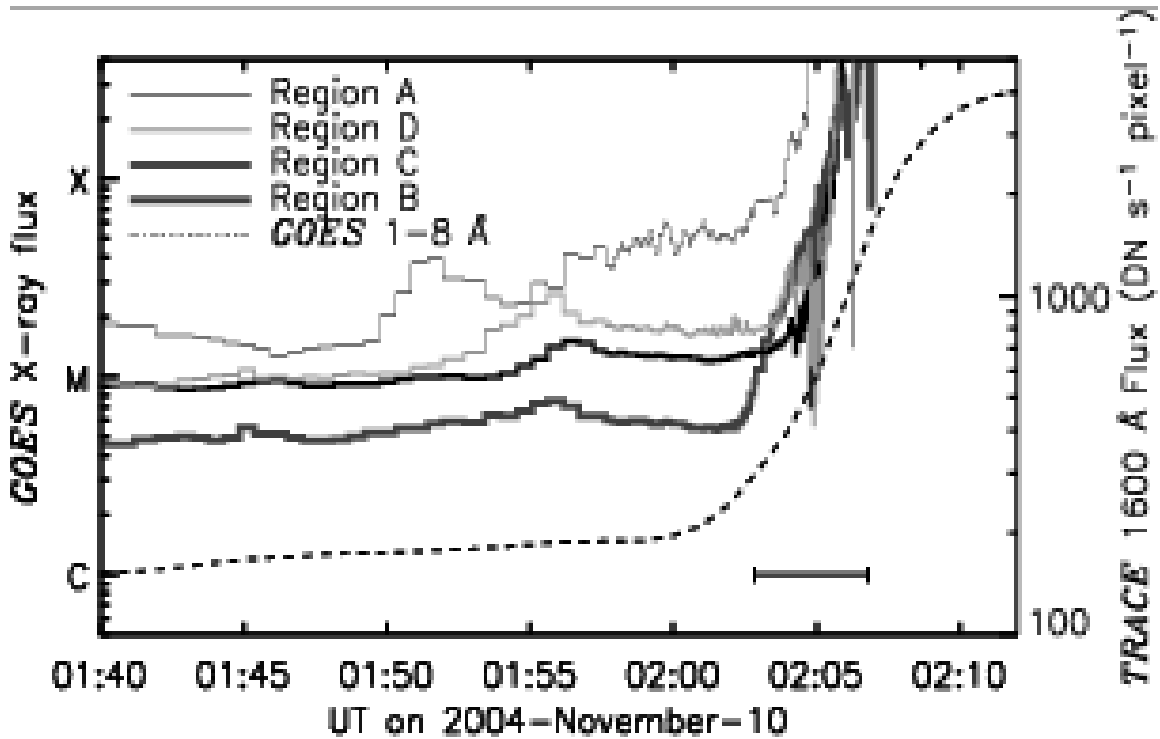}
\caption{Smooth dashed curve: {\sl GOES} soft X-ray flux; scale given in {\sl GOES} class on left vertical axis. Solid histogram curves: pixel-averaged {\sl TRACE} 1600-{\AA} intensity lightcurves from the four boxed regions shown in Fig.~\ref{traceimages}a); scale on right vertical axis. The horizontal bar indicates the range of time covered by data in Fig.~\ref{htprof}.\label{fourlcs}}
\end{center}
\end{figure}

\label{obserup}
At approximately 01:51:40~UT, a highly localized brightening occurred close to the IL at $A^+$ (Fig.~\ref{traceimages}a; see corresponding lightcurve in Fig.~\ref{fourlcs}). This brightening is approximately co-spatial with the emergence of a small bipole (hereafter indicated by $EF$), north of the IL (Fig.~\ref{traceimages}e); we assume that its lack of appearance until 03:15~UT is due both to MDI's long cadence and to the small scale of the bipole emerging in a strong, positive-flux environment.

The initial brightening at $A^+$ was followed by near-simultaneous brightenings at this and three other sites (marked $B^-$, $C^+$ and $D^-$ in Fig.~\ref{traceimages}a; superscripts indicate magnetic polarity) between 
$\sim$01:53~UT and 01:58~UT. These four, ribbon-like brightenings precede the later flare ribbons, the latter being %connect sites $A^+$ and $C^+$ on one side of the IL, and $B^-$ and $D^-$ on the other; these ribbons 
 detected as sharp rises in 1600-{\AA} flux from $\sim$02:02~UT (Fig.~\ref{fourlcs}).

The eastern leg of $f$ began to become visible in emission at 02:01:26~UT, with the full filament visible by 02:02:50~UT. The filament's curved axis had already developed a slight cusp shape near its apex by the latter time, which proceeded to develop further (Figs.~\ref{traceimages}c,d). The eruption of this filament is associated with the development of J-shaped  flare ribbons on both sides of the IL (Fig.~\ref{traceimages}c).

%@@@@@@@@@@@@@@@@@@@@@@@@@@@@@@@@@@@@@@@@@@@@@@@@@@
\section{Interpretation}
\label{physics}
\subsection{Eruption topology \& kinematics}
\label{topology}
\begin{figure}
\begin{center}
\includegraphics[width=5.0in]{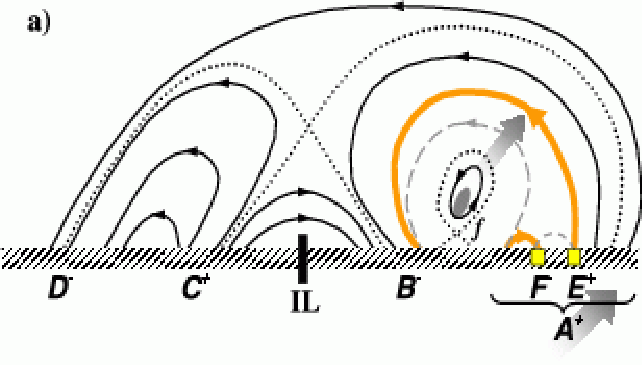}\\
\includegraphics[width=5.0in]{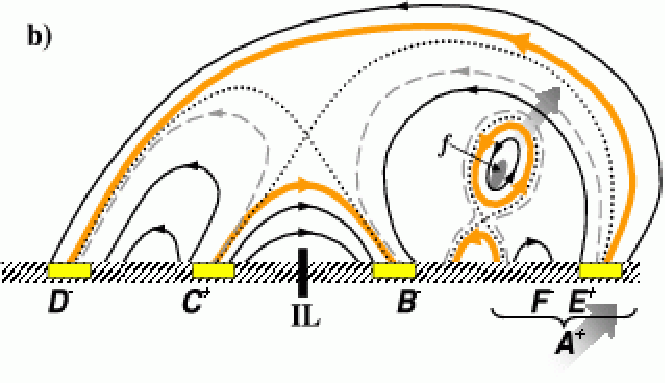}
%N.B. for online edition use the files f3a_colour.eps and f3b_colour.eps instead, as follows:
%\includegraphics[width=3.25in]{f3a_colour.eps}\\
%\includegraphics[width=3.25in]{f3b_colour.eps}
\caption{Main topological features of the magnetic configuration, with a simplified realization along the curved IL, folded into a 2D diagram. Dotted lines represent key separatrices; dashed lines, former connectivities later transformed by reconnection; continuous thick lines, new connectivities after reconnection; and continuous thin lines, continuing connectivites. a) Evolution in the initial phase of the eruption assuming a tether-weakening due to reconnection with the new flux, $EF$. b) The eruption with reconnection both above (quadrupolar) and below the erupting filament. The bottom-right arrow in each panel indicates the strongest photospheric motions parallel to the IL.\label{cartoons}}
\end{center}
\end{figure}

The initial brightening at $A^+$ indicates that the emerging bipole $EF$ (Sect.~\ref{obserup})  reconnects with the pre-existing coronal field to give new connectivities (Fig.~\ref{cartoons}a). The newly reconnected field lines above the filament $f$ are longer than the original ones, and so provide less downward stabilizing tension. Moreover, as with all emerging flux, the bipole $EF$ carries magnetic energy and helicity and a fraction is transferred to the large, reconnected loops, rendering them more stressed and thus able to expand further.
   
The four observed precursor ribbons are evidence that a quadrupolar-like reconnection occurred in this eruption. This may seem surprising in this magnetic configuration, formed mainly by a positive and a negative magnetic region, so similar to a bipolar magnetic arcade.  However, both positive and negative regions are divided into separate magnetic concentrations, and the IL is curved. This type of configuration is known to have a complex topology, not necessarily in the classical sense (i.e., with separatrices) but in the broader context of Quasi-Separatrix Layers (QSLs; \citealt{Demoulin97}).  QSLs are very thin layers where current sheets, and subsequently 3D magnetic reconnection, can develop.  The equivalent 2D magnetic topology is quadrupolar, as shown in Fig.~\ref{cartoons}.  The emergence of bipole $EF$ and -- on a longer time scale -- the observed shearing motions in the vicinity of $f$ (Sect.~\ref{obsall}) both contribute to an expansion of the magnetic configuration supporting $f$.  This can trigger quadrupolar reconnection, which weakens the tethers of the overlying arcade. The four ribbons, and plausibly also the release of $f$, are then interpreted naturally by the lateral breakout mechanism described in Sect.~\ref{Introduction} and Fig.~\ref{cartoons}b.

The erupting filament, however, shows evidence of a kinked shape as soon as it becomes observable by {\sl TRACE}. Moreover, the associated eruptive ribbons are J-shaped (Fig.~\ref{traceimages}c), just as expected with the QSLs computed for twisted magnetic configurations \citep*{Demoulin96b}. The separation of the ribbons along the IL, as well as their reversed-J shape, are characteristic of a negative helicity, again supporting the conclusion derived from the magnetic evolution (Sect.~\ref{obsall}). The main origin of the eruption is therefore most likely the accumulated magnetic stress, which progressively drives the configuration toward being kink-unstable.

In order to quantify the rise behaviour of the kinking filament with time, the projected position of its apex was determined by selecting co-ordinates at the apparent apex using a mouse cursor. We define the projected apex height as the distance, in the plane of the sky, between the apex and the point directly between the extremities of the filament as seen at 195~{\AA} (Fig~\ref{traceimages}b). The accuracy of these positions was tested by a series of independent measurements.  The errors in these measurements are estimated as the quadrature sum of {\sl TRACE}'s FWHM (0.5\arcsec) and the 1-pixel error on the screen during measurement, in each direction: this yields a 1$\!\sigma$ error in height of 1\arcsec. 

 We then fit these height-time data (Fig.~\ref{htprof}) with several growth laws
(constant acceleration, linearly increasing acceleration, power-law, exponential) 
and find that it is difficult to distinguish between these models from the limited data in {\sl TRACE}'s field of view. The most satisfactory fit is to an exponential function:
\begin{equation}
\label{fitfun}
h(t) = h_0 +  a_0\exp{(t/\tau)}
\end{equation}
where $(h_0+a_0) = 1.5\times10^4 \pm 6\times10^2$~km is the height of the filament at 02:02:50~UT ($t$=0), $a_0 = 6.7 \times 10^3 \pm 2.7\times 10^2$~km, and $\tau$ is the $e$-folding time (or `growth time') of the exponential acceleration term.
In equation~(\ref{fitfun}), $\tauÂ$ is unaltered by projection effects, and is found to be best fit by $\tau = 95 \pm 2$~s. The power-law and constant-acceleration models were both found to fit the data comparably well in terms of residuals, but these yielded start times near 02:04:15~UT; both the {\sl TRACE} and {\sl GOES} observations suggest, however, that the eruptive process had already started by 02:02:50~UT.

%@@@@@@@@@@@@@@@@@@@@@@@@@@@@@@@@@@@@@@@@@@@@@@@@@@
\subsection{Numerical simulation}
\label{numerique}

\begin{figure}
\begin{center}
\includegraphics[height=6.0in,angle=90]{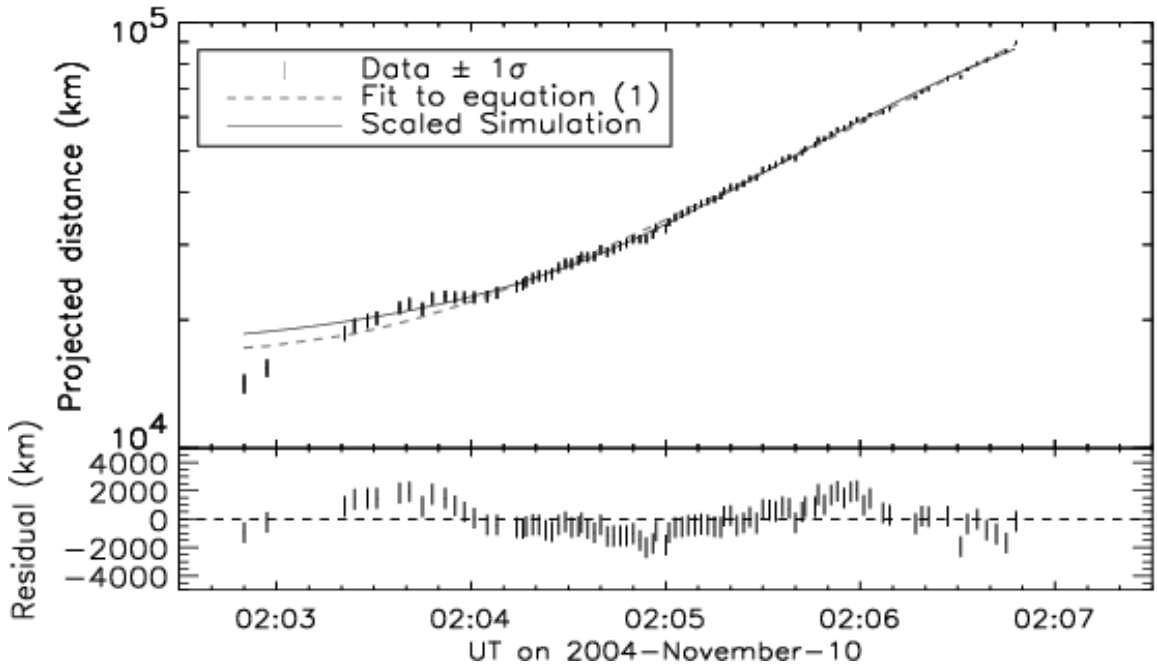}
\caption{Log-linear plot of the height-time profile of the filament apex in the plane of the sky. Vertical bars represent data points with $1\sigma$ errors.  The dashed line 
represents the best fit to the data by equation~(\ref{fitfun}). The solid line represents the $h(t)$ data of the simulation shown in Fig.~\ref{traceimages}. Lower panel shows residuals to the fit.\label{htprof}}
\end{center}
\end{figure}

The erupting filament $f$ exhibits a clear helical deformation of its axis shape (Fig.~\ref{traceimages}) as well as an apparently exponential growth in apex height (Fig.~\ref{htprof}). These observations suggest that the filament eruption itself is primarily driven by the KI of a magnetic flux rope which contains the filament.

To substantiate this assumption, we compare our observations with the numerical simulation of \cite{tk05}.  They used a modified version of the coronal flux rope equilibrium developed by \cite*{1999A&A...351..707T} as the initial condition in 3D, ideal-MHD simulations.  
By prescribing a supercritical flux-rope twist and a sufficiently weak ambient magnetic field, the kink-unstable flux rope showed an initially exponential rise, followed by a rise with a high constant velocity.

The Titov \& D\'emoulin equilibrium can be regarded as a model of a bipolar active region with the flux rope supporting a filament.  The large-scale photospheric flux distribution is qualitatively similar to the distribution in Region 10696 before the eruption: two main polarity regions are divided by a curved IL \cite[see Fig.~4 in][]{1999A&A...351..707T}.  The flux rope orientation in the model is nearly aligned with the IL (Figs.~\ref{traceimages}f thru \ref{traceimages}h), as is generally the case for filaments.

In Fig.~\ref{traceimages}, we compare the evolution of the axis shape 
of the filament with the simulation.  The point of view towards the model 
flux rope is chosen such that: i) the IL is oriented as in the observation; and ii) the morphology of the filament 
and the flux rope match as close as possible.  The (left-handed) 
helical axis deformation is qualitatively very similar on both counts. The path of the flux-rope apex deviates from radial ascent by about $\pi/4$ northward in projection, presumably in the direction of the weakest overlying field (as one can extrapolate from Fig.~\ref{traceimages}e, and as indicated in Fig.~\ref{cartoons})

By comparing the morphology at corresponding times, we also find that the observed helical deformation of the filament is initially stronger than in the simulation.  This indicates that the magnetic field strength, at least in the coronal region just above the filament, decreases more slowly with height than in the model (see Figs.~1 and 5 in \cite{tk05} for a comparison of the helical flux rope deformation in two configurations with different ambient fields). Future quantitative studies may enable the estimation of the height profile of the overlying magnetic field strength.

A scaling of the simulation data to the observed height-time profile, using an initial height of 15.5~Mm at 02:01:10~UT and an Alfv\'en time of 11.8~s ($v_A = 1300$~km~s$^{-1}$), also yields excellent agreement (Fig.~\ref{htprof}), further supporting the conclusion, from the fitting, that the rise was exponential in nature.

%@@@@@@@@@@@@@@@@@@@@@@@@@@@@@@@@@@@@@@@@@@@@@@@@@@

\section{Conclusions}
\label{Conclusions}

The emerging bipole north of the IL, spatially coincident with a pre-eruption brightening in {\sl TRACE}'s 1600~{\AA} filter, suggests an initiation role for tether-weakening in the release of an already highly twisted filament. However, this bipole is relatively small ($\Phi\approx 4.0\times10^{18}$~Mx) when compared with the flux of the active region ($\Phi\approx 4.5\times10^{22}$~Mx), and its role in weakening the field above the filament's flux rope is likely to be rather limited.  Its true contribution could rather be to bring a small amount of magnetic stress (or helicity), pushing the twisted configuration toward the critical value for eruption or triggering of the quadrupolar reconnection which leads to a strong tether weakening. Four bright ribbons are also observed before the filament eruption: they are the signatures of a quadrupolar reconnection taking place well above the filament, and so of a lateral breakout mechanism. In this case, as for \cite{2005ApJ...618.1001H}, evidence of both tether-weakening and breakout is present in the eruption.  Furthermore, the filament eruption itself appears to follow an exponentially growing form (cf.~\citealt{gallaghercme,2003SoPh..215..185S}), suggestive of an instability.  The apparent writhe of the filament, and hence of the host flux rope, points to the MHD helical kink instability as the likely driver behind the eruption. Moreover, the characteristics of the filament eruption have the same global characteristics as the MHD simulations of a kink-unstable magnetic flux rope.
 
We conclude that the release of the accumulated magnetic stress in the corona can involve several mechanisms, three of them being present in the studied event. The magnetic tethers are weakened both at low heights and well above the filament. However, the main engine of the filament eruption itself has all the observable characteristics of a kink instability. 
\acknowledgments

The authors are very grateful to Mitch Berger and Jingxiu Wang for helpful discussions, and to Jason King for valuable help in data retrieval. Work supported by grants OTKA T038013 \& OTKA T048961 (LVDG), DFG MA 1376/16-2 (BK) and computing time at NIC, J\"ulich.

%\bibliographystyle{apj2}
%
%\bibliography{david}

\clearpage

\clearpage

\clearpage

\clearpage

\end{document}